\documentclass[modern]{aastex62}

\usepackage{graphicx}
\usepackage{amsmath}
\usepackage{longtable}
\usepackage{color}
\usepackage{enumerate}

\graphicspath{{./}{figures/}}

\shorttitle{Black hole candidates with giant companions}
\shortauthors{Gu et al.}

\def\eff{\rm eff}
\def\dex{\rm dex}
\def\K{\rm K}
\def\L{\rm L}
\def\S/N{\rm S/N}
\def\km/s{\rm km~s^{-1}}

\begin{document}

\title{A Method to Search for Black Hole Candidates with Giant Companions \\
by LAMOST}

\correspondingauthor{Wei-Min Gu}
\email{guwm@xmu.edu.cn}

\author{Wei-Min Gu}
\affil{Department of Astronomy, Xiamen University, Xiamen,
Fujian 361005, P. R. China}
\author{Hui-Jun Mu}
\affiliation{Department of Astronomy, Xiamen University, Xiamen,
Fujian 361005, P. R. China}
\author{Jin-Bo Fu}
\affiliation{Department of Astronomy, Xiamen University, Xiamen,
Fujian 361005, P. R. China}
\author{Ling-Lin Zheng}
\affiliation{Department of Astronomy, Xiamen University, Xiamen,
Fujian 361005, P. R. China}
\author{Tuan Yi}
\affiliation{Department of Astronomy, Xiamen University, Xiamen,
Fujian 361005, P. R. China}
\author{Zhong-Rui Bai}
\affiliation{National Astronomical Observatories, Chinese Academy
of Sciences, Beijing 100012, P. R. China}
\author{Song Wang}
\affiliation{National Astronomical Observatories, Chinese Academy
of Sciences, Beijing 100012, P. R. China}
\author{Hao-Tong Zhang}
\affiliation{National Astronomical Observatories, Chinese Academy
of Sciences, Beijing 100012, P. R. China}
\author{Ya-Juan Lei}
\affiliation{National Astronomical Observatories, Chinese Academy
of Sciences, Beijing 100012, P. R. China}
\author{Yu Bai}
\affiliation{National Astronomical Observatories, Chinese Academy
of Sciences, Beijing 100012, P. R. China}
\author{Jianfeng Wu}
\affiliation{Department of Astronomy, Xiamen University, Xiamen,
Fujian 361005, P. R. China}
\author{Junfeng Wang}
\affiliation{Department of Astronomy, Xiamen University, Xiamen,
Fujian 361005, P. R. China}
\author{Jifeng Liu}
\affiliation{National Astronomical Observatories, Chinese Academy
of Sciences, Beijing 100012, P. R. China}
\affiliation{College of Astronomy and Space Sciences, University of
Chinese Academy of Sciences, Beijing 100049, P. R. China}

\begin{abstract}
We propose a method to search for stellar-mass black hole (BH) candidates
with giant companions from spectroscopic observations.
Based on the stellar spectra of LAMOST Data Release 6,
we obtain a sample of seven giants in binaries with
large radial velocity variation $\Delta V_R > 80~{\rm km~s^{-1}}$.
With the effective temperature,
surface gravity, and metallicity provided by LAMOST,
and the parallax given by {\it Gaia},
we can estimate the mass and radius of the giant,
and therefore evaluate the possible mass of the optically invisible star
in the binary.
We show that the sources in our sample are potential BH candidates,
and are worthy of dynamical measurement by further
spectroscopic observations.
Our method may be particularly valid for the selection of BH candidates
in binaries with unknown orbital periods.
\end{abstract}


\section{Introduction}\label{sec1}

According to the stellar evolution model, there may exist $\sim 10^8$
to $10^9$ stellar-mass black holes (BHs) in our Galaxy
\citep[e.g.,][]{Brown1994,Timmes1996}. However, only around 20 BHs
have been dynamically confirmed since the first BH was found in 1972
\citep{Bolton1972,Webster1972}. In addition, there are tens of BH
candidates without dynamical identification.
In total, the sum of confirmed BHs and candidates is less than
a hundred in our Galaxy \citep{Corral2016}.
It is known that most of the confirmed BHs and candidates were
originally selected from the X-ray observations. In general, an X-ray burst
in a binary means that there may exist a neutron star (NS) or a BH in
the binary. If there is no typical characteristic of NS systems
in the radiation, such as the type I X-ray burst or the radio pulses,
the compact object can be regarded as a BH candidate.
If the follow-up dynamical measurement
with spectroscopic observations
can derive that the compact object mass is larger than
$3 M_{\sun}$ \citep[e.g.,][]{Casares2014}, then a BH is identified.
Such a classic method,
based on the semi-amplitude of the radial velocity variation $K$
and the orbital period $P_{\rm orb}$ obtained from the radial velocity curve,
is well understood but may not be efficient.

The potential to search for BHs or BH candidates according to
some surveys have been widely studied.
The ability of {\it Gaia} on this issue has been investigated
\citep[e.g.,][]{Mashian2017,Breivik2017,Yamaguchi2018,
Yalinewich2018}, which predicted that hundreds or thousands of BHs
may be found by the end of its five-year mission.
\citet{Masuda2018} discussed the potential of the Transiting
Exoplanet Survey Satellite (TESS) to identify and characterize
nearby BHs with stellar companions on short-period orbits.
By exploring the Optical Gravitational Lensing Experiment in
its third generation (OGLE-III) database of 150 million
objects, \citet{Wyrzykowski2016} identified 13 microlensing
events which are consistent with having compact object lens.

A large number of BHs may exist in binaries, but without or with very
weak X-ray emission. Thus, different methods are required
in order to search for more existent BHs in our Galaxy.
For example, some physical parameters for binary systems
are well constrained by the spectroscopic observations
\citep[e.g.,][]{Mazeh1992,Marsh1994,Duemmler1997}.
The Large sky Area Multi-Object fiber Spectroscopic Telescope
\citep[LAMOST; also called Guoshoujing Telescope;][]{Wang1996,Su2004,Cui2012}
survey is a large-scale spectroscopic survey.
In our opinion, the released huge number of LAMOST stellar
spectra enable us to search for BH candidates
through a specific way, i.e., without the
X-ray bursts but simply from the spectroscopic observations.

The LAMOST Experiment for Galactic Understanding and Exploration
survey of Milky Way stellar structure has derived millions of stellar
spectra. Around 9 million stellar spectra have been released by LAMOST.
We can search for BH candidates in binaries based on these spectra.
This work focuses on the binary system containing a giant star.
The remainder is organized as follows.
The method is described in Section~\ref{sec2}.
Our sample and data analyses are shown in Section~\ref{sec3}.
The sources in our sample are investigated in Section~\ref{sec4}.
Conclusions and discussion are presented in Section~\ref{sec5}.

\section{Method}\label{sec2}

In a binary system with a compact object, the optically visible star is
denoted as $M_1$, and the compact object is denoted as $M_2$.
For simplicity, a circular orbit is assumed for the binary.
Then, a basic dynamical equation takes the form:
\begin{equation}
\frac{GM_2}{a^2} = \frac{V_{\rm K}^2}{a_1} \ ,
\label{VK}
\end{equation}
where $a$ is the separation of the binary, and $a_1$ is the distance between
$M_1$ and the center of mass, with $a_1/a = M_2/(M_1+M_2)$.
$V_{\rm K}$ is the Keplerian velocity of $M_1$ in the circular orbit.
In addition, we introduce a parameter ``$K$" to describe
the semi-amplitude of the radial velocity variation during a circle.
Thus, $K = V_{\rm K} \sin i$, where $i$ is the inclination angle of the
orbital plane.

We define $R_{\rm L1}$ as the effective Roche-lobe radius, which is
expressed as \citep{Eggleton1983}
\begin{equation}
\frac{R_{\rm L1}}{a} = \frac{0.49 q^{2/3}}{0.6 q^{2/3} + \ln (1+q^{1/3})} \,
\label{RL1}
\end{equation}
where $q \equiv M_1/M_2$.
By combining Equations (\ref{VK}-\ref{RL1}) we obtain
\begin{equation}
\frac{K^2 R_{\rm L1}}{G M_1 \sin ^2 i} = f(q)
= \frac{0.49 q^{-1/3} (1+q)^{-1}}{0.6 q^{2/3} + \ln (1+q^{1/3})} \ .
\label{fq}
\end{equation}
In general, we can assume $R_1 \leqslant R_{\rm L1}$.
If $M_1$ can just fill its Roche lobe, then we have
$R_1 = R_{\rm L1}$. Otherwise, the Roche lobe is not filled out,
thus $R_1 < R_{\rm L1}$. In this work,
we define $R_1 = \lambda R_{\rm L1}$ ($0 < \lambda \leqslant 1$).
The radius can be expressed as $R_1 = (G M_1/g)^{1/2}$, where $g$ is the
surface gravity. Then, Equation~(\ref{fq}) can be modified as
\begin{equation}
\frac{K^2}{\lambda \sin ^2 i \sqrt{G M_1 g}} = f(q) \ .
\label{fq1}
\end{equation}
Obviously, there exists $K \geqslant \Delta V_{R}/2$, where
$\Delta V_{R}$ is the largest variation of the radial velocity
among all the spectroscopic observations for a certain source.
Thus, Equation~(\ref{fq1}) implies
\begin{equation}
\frac{\Delta V_R^2}{4\lambda \sin ^2 i \sqrt{G M_1 g}} \leqslant f(q) \ .
\label{fq2}
\end{equation}
Interestingly, $f(q)$ is a monotonic function,
i.e., $f(q)$ always decreases with increasing $q$.
For a given pair of parameters ($i$, $\lambda$), as indicated by the above
inequality, with the known values for $\Delta V_R$, $\log g$, and $M_1$
from spectroscopic observations, the lower limit for $f(q)$
corresponds to an upper limit for $q$ (denoted as $q^{\rm max}$), thus
a lower limit for $M_2$ (denoted as $M_2^{\rm min}$).
For a reasonable pair ($i$, $\lambda$), if
$M_2^{\rm min} > 3 M_{\sun}$ and $M_2^{\rm min} > M_1$
are both matched
(since $M_1$ is already a giant and therefore $M_2^{\rm min} > M_1$ means
that $M_2$ cannot be a main sequence star),
the optically invisible star may be regarded as a potential BH candidate.
In particular, for the extreme case with $\sin i = 1$ and $\lambda = 1$,
if $M_2^{\rm min} > 3 M_{\sun}$ and $M_2^{\rm min} > M_1$ are both
satisfied, then the object $M_2$ can be regarded as an identified BH.

Here, we would stress that our method is particularly introduced for the case
that the orbital period of the binary is unknown. Otherwise,
as mentioned in Section~\ref{sec1},
$M_2$ can be estimated by the classic method,
which is based on the semi-amplitude $K$ and the orbital period
$P_{\rm orb}$.

\section{Sample and analyses}\label{sec3}

Recently, a huge number of stellar spectra from LAMOST Data Release 6
\footnote{\url{http://dr6.lamost.org/}\label{LDR6}} (hereafter LDR6),
together with the released observations of {\it Gaia} Data Release 2
(hereafter~GDR2, \citealp{Gaia2018}),
enable us to search for BH candidates through a specific way,
i.e., simply from the spectroscopic observations.
In LDR6, the ``A, F, G and K type star catalog" \textsuperscript{\ref{LDR6}}
provides the important stellar astrophysical parameters, such as
the effective temperature $T_{\eff}$, the surface gravity $\log g$,
the metallicity $[{\rm Fe/H}]$, and their errors by the LAMOST Stellar
Parameter pipeline (LASP, \citealp{Luo2015}).
In addition, the LAMOST 1D pipeline works on the measurement of heliocentric
radial velocity ($V_{R}$) for stars by using a cross-correlation method
\citep{Luo2015}.

\citet{Liu2015} selects the metal-rich giant stars with
$3500 <T_{\eff} < 6000~\K$, $\log g < 4.0~\dex$, and $[\rm Fe/H]> -0.6~\dex$.
In this work, the giant stars are selected by containing repeated radial
velocity measurements (at least three times) in LDR6 within $3^{\arcsec}$,
and match the following selection criteria:
\begin{enumerate}[(i)]
\item The signal-to-noise ratio ``$\rm S/N > 10$" is required
in the $g$ \footnote{We use the S/N at $g$ band because most of
the spectral lines sensitive to $\log g$ are located in
the range of wavelength ($\sim 4000 - 5300 \rm {\AA}$)
\citep{Liu2014,Liu2015}.} and $r$ bands \footnote{The radial velocity
from LAMOST is mainly from the lines in $\sim 4000-6600\rm {\AA}$.}.
\item Our main focus is the G and K giant stars.
Thus, the surface gravity $\log g < 3.0~\dex$ and the
effective temperature $3500~{\rm K} <T_{\eff} < 6000~{\rm K}$ are adopted.
\item The stellar spectra have only shifted single-line
(without double peaks) with a large variation of radial velocity,
$\Delta V_R > 80~\km/s$
\footnote{According to Inequality~(\ref{fq2}),
a larger $\Delta V_R$ corresponds to a higher $M_2^{\rm min}$, so we
choose a relatively large velocity ($80~\km/s$) as a lower limit, beyond
which there is a sample of several sources for detailed investigation.}.
\end{enumerate}

Our sample consists of seven binaries, each of which has a giant star.
The observational data of our sample are shown in Table~\ref{T1}.

\section{Black hole candidates}\label{sec4}

\begin{figure}
\plotone{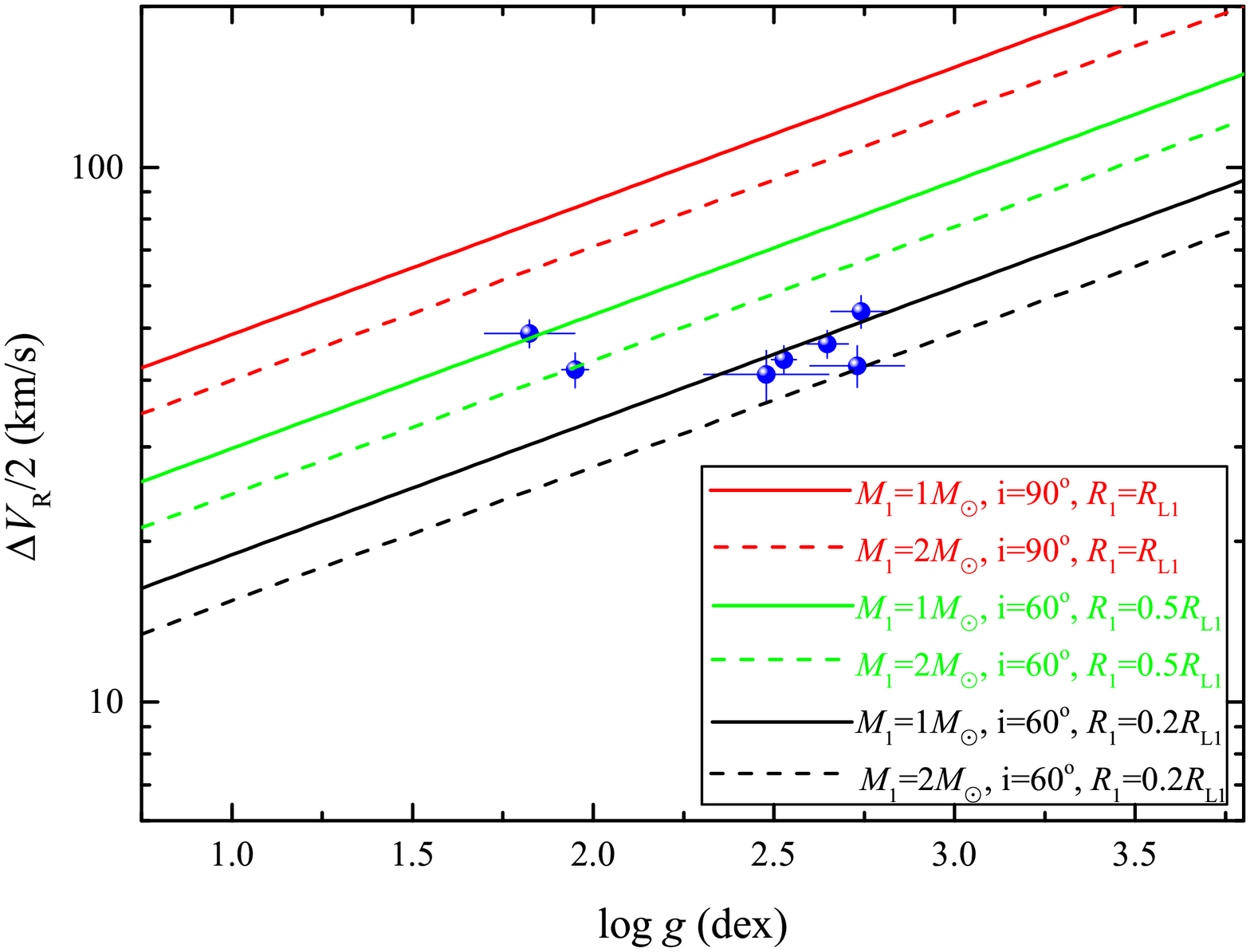}
\caption{Comparison of the observations (blue circles) with the theoretical
results (lines) in the $\Delta V_R/2 - \log g$ diagram,
where $M_2=3 M_{\sun}$ is adopted for the theoretical calculations.
\label{F1}}
\end{figure}

Following the method described in Section~\ref{sec2},
with a given pair of parameters ($i$, $\lambda$)
and the known values of $\Delta V_R$ and $\log g$ from LAMOST spectra,
we can study the possibility of BH candidates in our sample.
A comparison of the observations with the theory
in the $\Delta V_R/2 - \log g$ diagram is shown in Figure~\ref{F1}.
For the theoretical results, we choose the well-known critical mass
$M_2 = 3 M_{\sun}$ for the optically invisible star, and two typical
mass $M_1 = 1 M_{\sun}$ and $2 M_{\sun}$ for the observed giant star.
As indicated by Inequality~(\ref{fq2}), for a given $\log g$,
there is an upper limit for $\Delta V_R /2$.
If the observational $\Delta V_R /2$ is larger than the theoretical
maximal value, then a larger mass than $3 M_{\sun}$ is required for
$M_2$. In such case, the source is likely to be a BH candidate.
Here, we adopt the extreme case ($i=90^{\circ}$) and
a typical case ($i=60^{\circ}$) for the inclination angle.
In addition, for the radius ratio $\lambda = R_1/R_{\rm L1}$, we take
the extreme case ($\lambda =1$) and two reasonable cases
$\lambda = 0.5$ and 0.2
\footnote{As some prediction shows, e.g. Figure~1 of
\citet{Mashian2017}, a large fraction of BH binaries have
quite long orbital periods ($\sim$ years), where the radius of the optically
observed star can be far below the corresponding Roche-lobe radius.
Thus, the values of 0.5 and 0.2 for $\lambda$ are reasonable.}
for the analyses.
The theoretical results for $M_1 = 1 M_{\sun}$ and $2 M_{\sun}$
are respectively shown by the solid and dashed lines, where
the red, green and black lines correspond to
``$i=90^{\circ}$, $\lambda =1$", ``$i=60^{\circ}$, $\lambda = 0.5$",
and ``$i=60^{\circ}$, $\lambda = 0.2$", respectively.

All the seven sources in our sample are also plotted in this diagram
by the blue circles.
It is seen from Figure~\ref{F1} that, there is no source above
the red dashed or solid line, which means that
none of these sources can be regarded as an identified BH
according to the current spectroscopic observations.
All the sources are located between the green solid line and the
black dashed line,
which indicates that, for reasonable parameters such as $i\sim 60^{\circ}$
and $\lambda \sim 0.2-0.5$, the optically invisible star is likely to
be a BH candidate. Moreover, since the observational $\Delta V_R /2$ is
only a lower limit for the real $K$, the latter may be
significantly larger than the former,
particularly for the sources with only three times of observations.
If the physical parameter of the vertical axis in Figure~\ref{F1}
is replaced by the real $K$, the location of the seven sources should
be moved upwards.
Thus, more spectroscopic observations are required to make a judgement.

In addition, we check the seven sources in our sample to be real giant
stars through a different way, i.e., without using the values of $\log g$
from LAMOST.
The radius can be estimated by the relation
$L_{\rm bol} = 4\pi R_1^2 \sigma T_{\rm eff}^4$, where $L_{\rm bol}$ is the
bolometric luminosity of the giant star, and $T_{\rm eff}$ is given by LDR6.
In order to obtain the proper $L_{\rm bol}$, we should take the extinction
into account.
The reddening E($B-V$) is referred to the Pan-STARRS 3D Dust Map
\citep{Green2018}. The interstellar extinction $A_V$ is
calculated by using the Fitzpatrick reddening law: $R_V = 3.1$
\citep{Fitzpatrick1999}.
Then $L_{\rm bol}$ is calculated by the following:
(a) parallax given by {\it Gaia};
(b) $V$-band magnitude from UCAC4 \citep{Zacharias2012},
as shown in columns 4 and 5 of Table~\ref{T1};
(c) extinction $A_V$;
(d) Bolometric Correction as a function of $T_{\rm eff}$ \citep{Torres2010}.
The derived radius $R_1^{\rm LT}$ is shown in Table~\ref{T1},
where the superscript ``LT" means that the radius is based on the bolometric
luminosity $L_{\rm bol}$ and the effective temperature $T_{\rm eff}$.
Moreover, GDR2 provides the radius $R_1^{\rm G}$
for some sources in our sample, which is also shown in Table~\ref{T1}.
It is seen by $R_1^{\rm LT}$ and $R_1^{\rm G}$
that the sources in our sample are real giant stars with $R_1 \gg R_{\sun}$.

In order to evaluate the mass for the optically invisible star in
the binary, we first obtain a more convincing pair of ($M_1$, $R_1$)
through the PARSEC model \footnote{see the PARSEC model
\url{http://stev.oapd.inaf.it/cgi-bin/cmd_3.1} for details},
by given $T_{\eff}$, $\log g$, and [Fe/H] from LAMOST.
The obtained stellar parameters (the age, $R_1$, and $M_1$) are presented
in Table~\ref{T2}, which shows that $M_1$ is in the range $1\sim 2~M_{\sun}$.
That is why we choose $M_1 = 1 M_{\sun}$ and $2 M_{\sun}$ for the theoretical
analyses in Figure~\ref{F1}.

\begin{figure}
\plotone{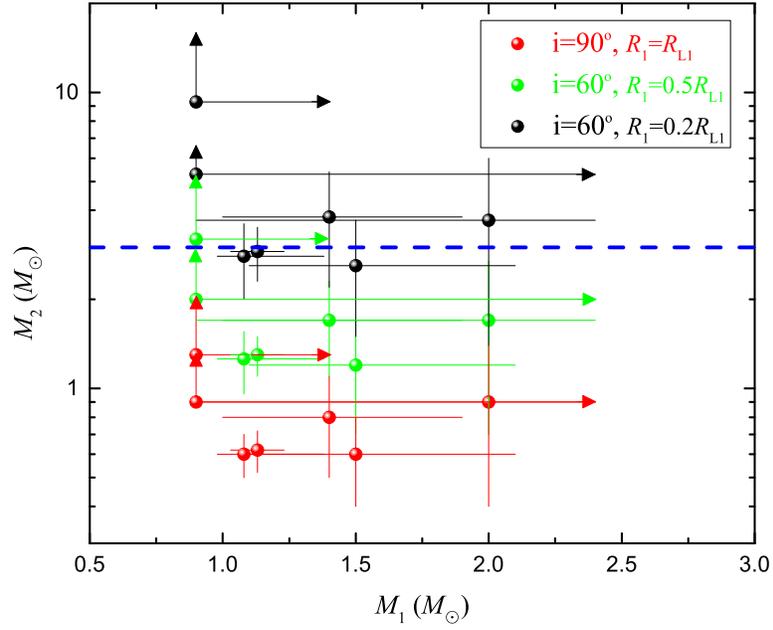}
\caption{Evaluation of $M_{2}$ for the seven sources in our sample
based on Equation~(\ref{fq3}),
where three pairs of ($i$,~$\lambda$) are adopted.
The blue dashed line represents the critical mass $M_{2}=3 M_{\sun}$.
\label{F2}}
\end{figure}

Based on the obtained $M_1$ and $R_1$ in Table~\ref{T2},
and the given pair of parameters ($i$,~$\lambda$),
together with the simple assumption $K= \Delta V_R /2$,
we can derive the mass $M_2$ by solving the following equation:
\begin{equation}
\frac{\Delta V_R^2 \cdot R_1}{4\lambda G M_1 \sin^2 i} = f(q) \ ,
\label{fq3}
\end{equation}
where the above equation is slightly transformed
from Equation~(\ref{fq}).
The results of our evaluation of $M_2$ for the seven sources in our sample
are shown in Table~\ref{T2} and Figure~\ref{F2}
\footnote{As shown in Table~\ref{T2}, two sources (No.1 and No.5)
have the same value for $M_1$, and have the same value for $M_2$ for a given
pair ($i$, $\lambda$), so the corresponding two circles should be located
at the same position for each color. Here, we plot with the derived values
without restriction to two significant digits, so that the two circles can be
distinguished.},
where three pairs of ($i$,~$\lambda$) are adopted.
For sources No.2 and No.7, as shown in Table~\ref{T2},
$M_1 = 0.9 M_{\sun}$ itself is the lower bound according to
the PARSEC model, below which the lifetime of the main sequence stage
is around or even exceeds the age of the Universe. Thus, arrows are
used instead of error bars in Figure~\ref{F2} for these two sources.
It is seen from Figure~\ref{F2} that, for the extreme case,
``$i=90^{\circ}$ and $\lambda =1$", the mass $M_2$ is below the critical
mass $M_2 = 3M_{\sun}$, which agrees with the results in Figure~\ref{F1}.
Thus, there is no identified BH in our sample according to the current
observations. However, for some reasonable parameters ``$i=60^{\circ}$
and $\lambda =0.5$ or $0.2$", Figure~\ref{F2} shows that the sources
can approach or even go beyond the critical blue dashed line, which
indicates that $M_2 > 3M_{\sun}$ is possible to be matched.
In addition, since there exists $M_1 \la 2M_{\sun}$ for our sample,
$M_2 > 3M_{\sun}$ means that $M_2 > M_1$ is simultaneously matched.
In such case, the optically invisible star is likely to be a BH candidate.
In our opinion, with regards to the observed large radial velocity
variation and the large radius of $M_1$, all the sources in our sample
have the potential to be BH candidates, and they are worthy of
further dynamical measurement.

Moreover, we check whether the sources in our sample
have been studied in published catalogs.
We cross-match our sources with the radio and X-ray catalogs using
a matching radius of $10\arcsec$.
No corresponding radio source was found according to the FIRST CATALOG
and the 1.4GHz NRAO VLA Sky Survey (NVSS).
Only one source (No.4) was known as a faint X-ray source according to
the ROSAT observations ($4.38 \arcsec$ of the distance to the center,
\citealp{Voges2000})
\footnote{We use HEASARC Browse to cross-match our sources
with the {\it Chandra}, XMM-Newton, {\it Swift}, and ROSAT observations.}.
Thus, source No.4 is likely a BH or an NS system with mass transfer
from the giant to the compact object.
Furthermore, source No.7 has measurements by Tycho-2 \citep{Hog2000}.

\section{Conclusions and discussion}\label{sec5}

In this work, we have proposed a method, from spectroscopic
observations by LAMOST, to search for stellar-mass BH candidates with
giant companions. Based on the spectra of LDR6, we have
derived a sample of seven giants in binaries with large radial velocity
variation $\Delta V_R > 80~{\rm km~s^{-1}}$.
With $T_{\rm eff}$, $\log g$, and [Fe/H] provided by LAMOST
and the parallax given by {\it Gaia}, we can estimate the values for
$M_1$ and $R_1$.
Moreover, we have evaluated the possible mass of $M_2$
for the extreme case $i=90^{\circ}$ and a typical case $i=60^{\circ}$,
and the extreme case $\lambda = 1$ and two reasonable
cases $\lambda = 0.5$ and $0.2$.
We argue that the sources in our sample are potential BH candidates
and worthy of further dynamical measurement.
Our method may be particularly valid to search for BH candidates
in binaries with unknown orbital periods.

Our analyses are based on the circular orbit assumption.
However, the binary system may have an elliptical orbit
(the eccentricity $e \neq 0$).
\citet{Moe2017} investigated the relation between the orbital
period and the eccentricity for early-type binaries identified by
spectroscopy, which showed that binaries have small
eccentricities $e \la 0.4$ for relatively short orbital periods
($P_{\rm orb} \la 20~{\rm days}$). On the other hand, for intermediate and
long periods, the range of eccentricity can be much wider, as shown by
their Equation~(3) and Figure~(6).
In the cases with eccentricity,
we can assume that the Roche-lobe radius at the pericenter should not
be less than $R_1$ (otherwise, if the optically invisible star
is an NS or a BH, strong X-ray radiation can be produced when the
giant star passes through the pericenter).
Thus, for a given $R_{\rm L1}$, the lower limit for the separation $a$
(major axis of the elliptical orbit) should be enhanced for $e>0$,
and therefore the lower limit $M_2^{\rm min}$ will be even larger.
For a rough estimate, we have the following analytic
inequality instead of Inequality~(\ref{fq2}) in Section~\ref{sec2}:
\begin{equation}
\frac{\Delta V_R^2 (1+e)}{4\lambda \sin ^2 i \sqrt{G M_1 g}} \leqslant f(q) \ ,
\end{equation}
which implies an even larger $M_2^{\rm min}$ than the case with $e=0$.
Thus, our analyses based on the circular orbit assumption are reliable,
in particular for the analytic lower mass limit for $M_2$.

The possibility of identifying BHs in detached binaries has been discussed
for decades \citep{Guseinov1966,Trimble1969},
where a method was provided based on single-line spectroscopic binaries
with large radial velocity variations.
With such a method, a BH candidate has recently been discovered
with an orbital period of 83~days \citep{Thompson2018}.
We should stress the difference between their method and ours.
In their method, the orbital period is a necessary condition for
the calculation.
On the contrary, our method is based on the comparison of $R_1$ with
the Roche-lobe radius $R_{\rm L1}$, whereas the orbital period is unknown,
as discussed in the last paragraph of Section~\ref{sec2}.
In addition, based on near-infrared observations with the
VISTA Variables in the V\'{\i}a L\'{a}ctea Survey (VVV), a
microlensing stellar-mass BH candidate was discovered, which is likely
a good isolated BH candidate \citep{Minniti2015}.
Moreover, \citet{Giesers2018} found a detached stellar-mass BH candidate
in the globular cluster NGC~3201 by performing multiple epoch spectroscopic
observations.

Furthermore, we can expect that, with growing data
released by LAMOST, particularly for the ongoing
LAMOST Medium Resolution Survey,
the proposed method in the present work may be helpful to select
a large number of BH candidates with giant companions.
Moreover, our method can be directly applied to the SDSS APOGEE
data to search for BH candidates, which provide high-resolution spectra
on giant stars.

\acknowledgments

We thank Yi-Ze Dong, Fan Yang, and Mou-Yuan Sun for beneficial discussions,
and thank the referee for helpful suggestions that improved the manuscript.
This work was supported by the National Natural Science Foundation of China
(NSFC) under grants 11573023, 11603035, U1831205, 11425313 and 11333004.
Guoshoujing Telescope (the Large Sky Area Multi-Object Fiber Spectroscopic
Telescope, LAMOST) is a National Major Scientific Project built by
the Chinese Academy of Sciences. Funding for the project has been provided
by the National Development and Reform Commission. LAMOST is operated
and managed by the National Astronomical Observatories, Chinese Academy
of Sciences.

\clearpage

\clearpage

\clearpage

\begin{deluxetable*}{cllcccccccccc}
\tablecaption{Parameters for the sources in our sample. \label{T1}}
\tablewidth{1000pt}
\tablehead{
\colhead{No.}
&\colhead{RA}
&\colhead{DEC}
&\colhead{UCAC4}
& \colhead{Vmag}
& \colhead{$T_{\eff}$}
& \colhead{$\log g$}
& \colhead{$\rm [Fe/H]$}
&\colhead{$N_{\rm obs}$}
& \colhead{$\Delta V_R$}
& \colhead{$\varpi$}
& \colhead{$R_1^{\rm G}$}
& \colhead{$R_1^{\rm LT}$}\\
\colhead{}& \colhead{}& \colhead{}&\colhead{}& \colhead{(mag)}
& \colhead{($\K$)}& \colhead{($\dex$)}
& \colhead{($\dex$)}&\colhead{}& \colhead{($\km/s$)}& \colhead{($\rm mas$)}
& \colhead{($R_{\sun}$)}& \colhead{($R_{\sun}$)}\\
\colhead{(1)}& \colhead{(2)}& \colhead{(3)}& \colhead{(4)}
& \colhead{(5)}& \colhead{(6)}& \colhead{(7)}
& \colhead{(8)}& \colhead{(9)}& \colhead{(10)}& \colhead{(11)}
&\colhead{(12)}& \colhead{(13)}
}
\centering
\tablewidth{0pt}
\tabletypesize{\tiny}
\tablecolumns{13}
\tablenum{1}
\startdata
1	&	0.839246	&	38.518578	&	643-000232	&	12.736	$\pm$	0.05	&	4696 	$\pm$	36 	&	2.65 	$\pm$	0.06 	&	-0.25 	$\pm$	0.03 	&	3	&	93.5 	$\pm$	5.6 	&	0.505 	$\pm$	0.043 	&	9.15	&	10.8 	$\pm$	0.2 	\\
2	&	3.887134	&	38.688854	&	644-000944	&	12.665	$\pm$	0.09	&	4301 	$\pm$	22 	&	1.95 	$\pm$	0.04 	&	-0.47 	$\pm$	0.02 	&	4	&	83.7 	$\pm$	6.2 	&	0.379 	$\pm$	0.032 	&	15.82	&	19.7 	$\pm$	0.3 	\\
3	&	58.561194	&	45.43539	&	678-024276	&	15.054	$\pm$	0.03	&	4943 	$\pm$	108 	&	2.48 	$\pm$	0.17 	&	-0.65 	$\pm$	0.10 	&	3	&	81.9 	$\pm$	9.0 	&	0.148 	$\pm$	0.034 	&	--	&	14.7 	$\pm$	0.7 	\\
4	&	74.0532211	&	54.0059047	&	721-037069	&	12.784	$\pm$	0.11	&	4823 	$\pm$	53 	&	2.74 	$\pm$	0.08 	&	-0.29 	$\pm$	0.05 	&	4	&	107.5 	$\pm$	7.4 	&	1.068 	$\pm$	0.034 	&	6.34	&	7.0 	$\pm$	0.2 	\\
5	&	106.45855	&	13.606334	&	519-037128	&	14.51	$\pm$	0.02	&	4655 	$\pm$	21 	&	2.53 	$\pm$	0.03 	&	-0.31 	$\pm$	0.02 	&	3	&	87.5 	$\pm$	5.3 	&	0.122 	$\pm$	0.030 	&	--	&	18.4 	$\pm$	0.2 	\\
6	&	111.33637	&	28.067468	&	591-041200	&	14.698	$\pm$	0.07	&	4832 	$\pm$	83 	&	2.73 	$\pm$	0.13 	&	-0.23 	$\pm$	0.08 	&	6	&	85.1 	$\pm$	7.6 	&	0.152 	$\pm$	0.032 	&	--	&	11.9 	$\pm$	0.4 	\\
7	&	169.1288227	&	55.7284139	&	729-048720	&	10.638	$\pm$	0.17	&	4191 	$\pm$	80 	&	1.82 	$\pm$	0.12 	&	-0.75 	$\pm$	0.07 	&	3	&	97.8 	$\pm$	5.8 	&	1.086 	$\pm$	0.031 	&	13.14	&	17.9 	$\pm$	0.7 	\\
\enddata
\tablecomments{
Column (1): Number of the source.
Column (2): Right ascension (J2000).
Column (3): Declination (J2000).
Column (4): UCAC4 recommended identifier.
Column (5): V-band magnitude from UCAC4.
Column (6): Effective temperature from LDR6.
Column (7): Surface gravity from LDR6.
Column (8): Metallicity from LDR6.
Column (9): Times of observations.
Column (10): Observed largest variation of radial velocity.
Column (11): Absolute stellar parallax from GDR2.
Column (12): Radius of the giant star from GDR2.
Column (13): Radius of the giant star based on the relation of
the bolometric luminosity and the effective temperature.}
\end{deluxetable*}

\clearpage

\begin{deluxetable*}{lllcccc}
\tablecaption{Evaluation of $M_{2}$ for the sources in our sample. \label{T2}}
\tablewidth{1000pt}
\tablehead{
\colhead{No.}
&\colhead{Age}
&\colhead{$M_{1}$}
& \colhead{$R_{1}$}
& \colhead{$M_{2}^{\rm I}$}
& \colhead{$M_{2}^{\rm II}$}
&\colhead{$M_{2}^{\rm III}$}
\\
\colhead{}& \colhead{($10^9 {\rm yr}$)}& \colhead{($M_{\sun}$)}
& \colhead{($R_{\sun}$)}& \colhead{($M_{\sun}$)}
& \colhead{($M_{\sun}$)}& \colhead{($M_{\sun}$)}\\
\colhead{(1)}& \colhead{(2)}& \colhead{(3)}& \colhead{(4)}
& \colhead{(5)}& \colhead{(6)}& \colhead{(7)}
}
\centering
\tablewidth{0pt}
\tabletypesize{\scriptsize}
\tablecolumns{7}
\tablenum{2}
\startdata
1       &$      7.3^{+4.7}_{-3.8}$&$    1.1     ^{      +0.3    }_{     -0.1    }$&$    8.1     ^{      +1.5    }_{     -0.8    }$&     0.6     $\pm$   0.1     &       1.3     $\pm$   0.3     &       2.9     $\pm$   0.8     \\
$2^{{\dag}}$    &$      13^{    +2.2    }                       _{      -0.1    }       $&      $0.9 ^{ +1.5    }_{     -}      $&$     17.8    ^{      +0.6    }_{     -       }$&     $0.9^{+0.4}         _{      -       }$                             &       $2.0 ^{+0.7}_{     -       }$                             &       $5.3^{+0.7}  _{      -       }$         \\
3       &$      0.97    ^{      +9.64   }_{     -0.33   }       $&$     2.0     ^{      +0.4    }_{     -1.1    }$&$    13.3    ^{      +2.7    }_{     -3.7    }$&     0.9     $\pm$   0.5     &       1.7     $\pm$   1.0     &       3.7     $\pm$   2.3     \\
4       &$      3.2     ^{      +7.6    }_{     -1.9    }       $&$     1.4     ^{      +0.5    }_{     -0.4    }$&$    8.2     ^{      +2.2    }_{     -1.7    }$&     0.8     $\pm$   0.3     &       1.7     $\pm$   0.6     &       3.8     $\pm$   1.6     \\
5       &$      6.3     ^{      +3.1    }_{     -1.9    }       $&$     1.1     ^{      +0.1    }_{     -0.1    }$&$    9.5     ^{      +0.8    }_{     -0.6    }$&     0.6     $\pm$   0.1     &       1.3     $\pm$   0.2     &       2.9     $\pm$   0.6     \\
6       &$      2.6     ^{      +4.4    }_{     -1.6    }       $&$     1.5     ^{      +0.6    }_{     -0.4    }$&$    8.7     ^{      +2.8    }_{     -1.9    }$&     0.6     $\pm$   0.2     &       1.2     $\pm$   0.5     &       2.6     $\pm$   1.1     \\
$7^{{\dag}}$    &$      11      ^{      +1      }_{     -9      }       $&$     0.9 ^{  +0.5    }_{     -       }               $&$     20.8    ^{      +8.1    }_{     -       }$&     $1.3^{      +0.7    }_{     -       }$             &       $3.2^{      +1.8    }_{     -       }$                     &       $9.3^{      +5.2    }_{     -       }$             \\
\enddata

\tablecomments{
$^{{\dag}}$ The parameters of the second and the seventh sources cannot
be well constrained, where $M_1 = 0.9 M_{\sun}$ itself is
the lower bound according to the PARSEC model.
Column (1): Number of the source.
Column (2): Age of the giant star from the PARSEC model.
Column (3): Mass of the giant star from the PARSEC model.
Column (4): Radius of the giant star from the PARSEC model.
All the errors about the PARSEC model are 90\% confidence.
Column (5): Mass of $M_2$ for ``$i=90^{\circ}$, $\lambda = 1$".
Column (6): Mass of $M_2$ for ``$i=60^{\circ}$, $\lambda = 0.5$".
Column (7): Mass of $M_2$ for ``$i=60^{\circ}$, $\lambda = 0.2$".
}
\end{deluxetable*}


\begin{thebibliography}{}

\bibitem[Bolton(1972)]{Bolton1972}
Bolton, C.~T.\ 1972, \nat, 235, 271

\bibitem[Breivik et al.(2017)]{Breivik2017}
Breivik, K., Chatterjee, S., \& Larson, S.~L.\ 2017, \apjl, 850, L13

\bibitem[Brown \& Bethe(1994)]{Brown1994}
Brown, G.~E., \& Bethe, H.~A.\ 1994, \apj, 423, 659

\bibitem[Casares \& Jonker(2014)]{Casares2014}
Casares, J., \& Jonker, P.~G.\ 2014, \ssr, 183, 223

\bibitem[Corral-Santana et al.(2016)]{Corral2016}
Corral-Santana, J.~M., Casares, J., Mu{\~n}oz-Darias, T., et al.\ 2016,
\aap, 587, A61

\bibitem[Cui et al.(2012)]{Cui2012}
Cui, X.-Q., Zhao, Y.-H., Chu, Y.-Q., et al.\ 2012, Research in Astronomy and Astrophysics, 12, 1197

\bibitem[Drake et al.(2014)]{Drake2014}
Drake, A.~J., Graham, M.~J., Djorgovski, S.~G., et al.\ 2014, \apjs, 213, 9

\bibitem[Duemmler et al.(1997)]{Duemmler1997}
Duemmler, R., Ilyin, I.~V., \& Tuominen, I.\ 1997, \aaps, 123, 209

\bibitem[Eggleton(1983)]{Eggleton1983}
Eggleton, P.~P.\ 1983, \apj, 268, 368

\bibitem[Fitzpatrick(1999)]{Fitzpatrick1999}
Fitzpatrick, E.~L.\ 1999, \pasp, 111, 63

\bibitem[Gaia Collaboration et al.(2018)]{Gaia2018}
Gaia Collaboration, Brown, A.~G.~A., Vallenari, A., et al.\ 2018, \aap, 616, A1

\bibitem[Giesers et al.(2018)]{Giesers2018}
Giesers, B., Dreizler, S., Husser, T.-O., et al.\ 2018, \mnras, 475, L15

\bibitem[Green et al.(2018)]{Green2018}
Green, G.~M., Schlafly, E.~F., Finkbeiner, D., et al.\ 2018, \mnras, 478, 651

\bibitem[Guseinov \& Zel'dovich(1966)]{Guseinov1966}
Guseinov, O.~K., \& Zel'dovich, Y.~B.\ 1966, \sovast, 10, 251

\bibitem[H{\o}g et al.(2000)]{Hog2000}
H{\o}g, E., Fabricius, C., Makarov, V.~V., et al.\ 2000, \aap, 355, L27

\bibitem[Liu et al.(2014)]{Liu2014}
Liu, C., Deng, L.-C., Carlin, J.~L., et al.\ 2014, \apj, 790, 110

\bibitem[Liu et al.(2015)]{Liu2015}
Liu, C., Fang, M., Wu, Y., et al.\ 2015, \apj, 807, 4

\bibitem[Luo et al.(2015)]{Luo2015}
Luo, A.-L., Zhao, Y.-H., Zhao, G., et al.\ 2015,
Research in Astronomy and Astrophysics, 15, 1095

\bibitem[Marsh et al.(1994)]{Marsh1994}
Marsh, T.~R., Robinson, E.~L., \& Wood, J.~H.\ 1994, \mnras, 266, 137

\bibitem[Mashian \& Loeb(2017)]{Mashian2017}
Mashian, N., \& Loeb, A.\ 2017, \mnras, 470, 2611

\bibitem[Masuda \& Hotokezaka(2018)]
{Masuda2018} Masuda, K., \& Hotokezaka, K.\ 2018, arXiv:1808.10856

\bibitem[Mazeh \& Goldberg(1992)]{Mazeh1992}
Mazeh, T., \& Goldberg, D.\ 1992, \apj, 394, 592

\bibitem[Minniti et al.(2015)]{Minniti2015}
Minniti, D., Contreras Ramos, R., Alonso-Garc{\'{\i}}a, J., et al.\ 2015, \apjl, 810, L20

\bibitem[Moe \& Di Stefano(2017)]{Moe2017}
Moe, M., \& Di Stefano, R.\ 2017, \apjs, 230, 15

\bibitem[Paczy{\'n}ski(1971)]{Paczynski1971}
Paczy{\'n}ski, B.\ 1971, \araa, 9, 183

\bibitem[Remillard \& McClintock(2006)]{RM2006}
Remillard, R.~A., \& McClintock, J.~E.\ 2006, \araa, 44, 49

\bibitem[Su \& Cui(2004)]{Su2004}
Su, D.-Q., \& Cui, X.-Q.\ 2004, \cjaa, 4, 1

\bibitem[Thompson et al.(2018)]{Thompson2018}
Thompson, T.~A., Kochanek, C.~S., Stanek, K.~Z., et al.\ 2018, arXiv:1806.02751

\bibitem[Timmes et al.(1996)]{Timmes1996}
Timmes, F.~X., Woosley, S.~E., \& Weaver, T.~A.\ 1996, \apj, 457, 834

\bibitem[Torres(2010)]{Torres2010}
Torres, G.\ 2010, \aj, 140, 1158

\bibitem[Trimble \& Thorne(1969)]{Trimble1969}
Trimble, V.~L., \& Thorne, K.~S.\ 1969, \apj, 156, 1013

\bibitem[Voges et al.(2000)]{Voges2000}
Voges, W., Aschenbach, B., Boller, T., et al.\ 2000, \iaucirc, 7432, 1

\bibitem[Wang et al.(1996)]{Wang1996}
Wang, S.-G., Su, D.-Q., Chu, Y.-Q., Cui, X., \& Wang, Y.-N.\ 1996, \ao, 35, 5155

\bibitem[Webster \& Murdin(1972)]{Webster1972}
Webster, B.~L., \& Murdin, P.\ 1972, \nat, 235, 37

\bibitem[Wyrzykowski et al.(2016)]{Wyrzykowski2016}
Wyrzykowski, {\L}., Kostrzewa-Rutkowska, Z., Skowron, J., et al.\ 2016, \mnras, 458, 3012

\bibitem[Yalinewich et al.(2018)]{Yalinewich2018}
Yalinewich, A., Beniamini, P., Hotokezaka, K., \& Zhu, W.\ 2018, \mnras, 481, 930

\bibitem[Yamaguchi et al.(2018)]{Yamaguchi2018}
Yamaguchi, M.~S., Kawanaka, N., Bulik, T., \& Piran, T.\ 2018, \apj, 861, 21

\bibitem[Zacharias et al.(2012)]{Zacharias2012}
Zacharias, N., Finch, C.~T., Girard, T.~M., et al.\ 2012,
VizieR Online Data Catalog, 1322

\end{thebibliography}
\end{document}